\documentstyle{article}        

\begin{document}

\title{Exact perturbations for inflation with smooth exit}

\author{
Roy Maartens and Daniel Tilley
\\
\\
{\footnotesize School of Computer Science and Mathematics, Portsmouth
University, Portsmouth~PO1~2EG, Britain}
}

\maketitle

\begin{abstract}

Toy models for the Hubble rate or the scalar field potential
have been used to analyze the amplification
of scalar perturbations through a smooth transition from inflation to
the radiation era. We use a Hubble rate that arises
consistently from a decaying vacuum cosmology, which evolves
smoothly from nearly de Sitter inflation to radiation domination.
We find exact solutions for super-horizon perturbations (scalar
and tensor), and
for sub-horizon perturbations in the vacuum- and radiation-dominated
eras. The standard conserved quantity for super-horizon scalar
perturbations is exactly constant for growing modes, and
zero for the decaying modes. \\

\end{abstract}


\section{Introduction}

The transition from inflation to radiation-domination involves
a number of subtle issues affecting the evolution of perturbations.
Recently, the controversy initiated by \cite{g2}
over whether indeed super-horizon
scalar perturbations are strongly amplified in inflation, appears
to have been settled, the answer being affirmative (see \cite{dm}
and the subsequent papers \cite{M1,g,ms}). The transition is
often approximated as an instantaneous jump, which is reasonable,
since super-horizon modes change on a timescale that is much
greater than the transition time. In order to avoid the
complications involved in the matching conditions for a jump
transition, smooth transitions have also been considered,
using toy models for the Hubble rate \cite{M1,ms} or the
potential of the scalar field \cite{g}.

Here we also consider a cosmology with a smooth
exit from inflation, but instead
of an ad hoc toy model, we use a recently proposed decaying vacuum
cosmology \cite{M9}. In this model,
a simple exact form for the Hubble rate
is deduced consistently from simple physical conditions. The
model, together with a brief summary of the necessary results
from perturbation theory, is discussed in Section 2.
Remarkably, the Hubble rate for the model, 
with the associated total energy
density and pressure, leads to an exact solution for super-horizon
scalar perturbations, which we present in Section 3.

The toy-model solutions of \cite{M1,g,ms} are given approximately or
numerically. The advantage of an exact solution lies in the
additional clarity and the definiteness with which certain questions
may be answered. As an example, we show that for our
solution, the standard
conserved quantity for growing modes is exactly constant, and
exactly zero for the decaying modes.
In line with the toy-model results, our solution also confirms the
standard picture of strong amplification of perturbations through
the transition from inflation to the radiation era.
We extend the toy-model results of \cite{M1,g} in
section 3 by finding the exact form of
super-horizon tensor perturbations.
For completeness, we also give
exact expressions for
scalar and tensor sub-horizon perturbations in the
vacuum- and radiation-dominated eras. 
In Section 4 we consider the extension of our results
to incorporate a second smooth transition from radiation to matter
domination.

\section{The inflationary model}

In a decaying vacuum cosmology (see e.g. \cite{l} and references
cited there),
the false vacuum, with energy density $\Lambda(t)$, 
decays into radiation, with energy density $\rho(t)$. The total
(conserved) energy density is $\rho_{\rm T}=\rho+\Lambda$, and the
total
pressure is $p_{\rm T}=
{1\over3}\rho-\Lambda$. These appear in the field
equations for a flat
Friedmann-Lemaitre-Robertson-Walker (FLRW) universe:
\begin{equation}
\rho_{\rm T} =3H^2\,,~~p_{\rm T}=-2\dot{H}-3H^2\,,
\label{3}\end{equation}
where $H=\dot{a}/a$ is the Hubble rate and $a$ is the scale factor.
The decay of the vacuum into
radiation is a non-adiabatic process, generating entropy and driving
inflation until the vacuum energy falls low enough for
inflationary expansion to end, and radiation domination to develop.
The transition is inherently smooth.
Despite the non-adiabatic interaction between the vacuum and created
radiation, the combined vacuum-radiation system behaves like a perfect
fluid, and standard adiabatic perturbation results may be applied to
analyse metric scalar and tensor fluctuations. 
This feature also applies to the more general class of `warm' 
inflationary models \cite{b}.

In \cite{M9}, a decaying vacuum model is subject to the following
simple physical conditions:\\
(a) the initial vacuum for radiation is
a regular Minkowski vacuum (i.e. with zero energy density, particle
number, entropy, etc.);\\
(b) the created radiation obeys the first law of
thermodynamics for open systems, i.e.
\[
d(\rho V)+p dV-\left({\rho+p\over n}\right)d(nV)=0 \,,
\]
where $p={1\over3}\rho$ is the radiation pressure, $n$ its number
density, and $V$ the comoving volume of the observable (causally
connected) universe; \\
(c) the total number of radiation particles
created throughout the expansion of the observable universe 
is finite.

It is then shown in \cite{M9} that these conditions strongly
constrain the expansion, and that the simplest
evolution consistent with the conditions is given by
a Hubble rate
\begin{equation}
H(a)=2H_{\rm e}\left({a_{\rm e}^2\over a^{2}+a_{\rm e}^2}\right)\,.
\label{2}
\end{equation}
Here $a_{\rm e}$ is 
the epoch of exit from inflation, 
defined by $\ddot{a}_{\rm e}=0$,
and $H_{\rm e}$ is  
the Hubble rate at exit. 
For 
$a\ll a_{\rm e}$, equation (\ref{2}) shows that
$H\approx$ constant, so that the initial evolution of the 
universe is
approximately de Sitter inflation.
For $a\gg a_{\rm e}$ the Hubble rate falls off as
$a^{-2}$, so that the universe becomes radiation-dominated. 
In this era, $\Lambda$ is negligible, falling off as $a^{-6}$.

The simple form (\ref{2}) for $H(a)$ had been introduced in
\cite{M2}, but as an ad hoc toy model, without a consistent
physical foundation as given in \cite{M9}.
The cosmic proper time follows on integrating equation 
(\ref{2}), as in \cite{M2}:
\begin{equation}
t=t_{\rm e}+{1\over 4H_{\rm e}}\left[
\ln\left({a\over a_{\rm e}}\right)^{2}+\left(a
\over a_{\rm e}\right)^{2}-1\right]\,,
\label{2a}\end{equation}
while equation (\ref{3}) gives
\begin{eqnarray}
\rho_{\rm T} &=& 12H_{\rm e}^{2}\left[{a_{\rm e}^{2}\over 
a^{2}+a_{\rm e}^2}\right]^2 \,, \label{6} \\
p_{\rm T} &=& 4H_{\rm e}^{2}\left[{a_{\rm e}^{4}(a^{2}-3a_{\rm e}^2)
\over (a^{2}+a_{\rm e}^2)^{3}}\right]\,.\label{5}
\end{eqnarray}
It follows that the effective pressure index and adiabatic sound
speed are given by
\begin{eqnarray}
w &\equiv& {p_{\rm T}\over \rho_{\rm T}}
={1\over 3}\left({a^{2}-3a_{\rm e}^2
\over a^{2}+a_{\rm e}^2}\right)\,,\label{7} \\
c_{\rm s}^2 &\equiv &{\dot{p}_{\rm T}\over \dot{\rho}_{\rm T}}=
{1\over 3}\left({a^{2}-5a_{\rm e}^2
\over a^{2}+a_{\rm e}^2}\right)\,.\label{8}
\end{eqnarray}

A physical length scale $\lambda$, corresponding to
a comoving wave number $k$, is given by
$\lambda=2\pi a/k$. 
This scale crosses the Hubble radius
$H^{-1}$ when $a=k/(2\pi H)$.
By equation (\ref{2}), the epoch $a_-$ of leaving and the epoch $a_+$
of re-entering are given exactly by
\begin{equation}
{a_\pm\over a_{\rm e}}={k_{\rm e}\over k}
\left[1\pm\sqrt{1-\left({k\over k_{\rm e}}\right)^2}\,
\right] \,.
\label{b}\end{equation}
where $k_{\rm e}=2\pi a_{\rm e}H_{\rm e}$ is the comoving wave number
of the Hubble radius at exit.
Scales with $k\geq k_{\rm e}$ do not cross the Hubble
radius, and remain sub-horizon.
Scales which leave the horizon well before
the end of inflation have $k\ll k_{\rm e}$. It follows from
(\ref{b}) that for these super-horizon scales
\[
{a_\pm\over a_{\rm e}}\approx \left({2k_{\rm e}\over k}\right)^{\pm 1}
\,.
\]
The fact that all
$k<k_{\rm e}$ modes re-enter during radiation domination
is a consequence of the simplistic nature of the model. A more
complete model requires a further transition from radiation to
matter domination, and this adjustment will invalidate the expressions
for $a_{+}$.

The evolution of Bardeen's gauge invariant potential $\Phi$
describing adiabatic scalar perturbations is 
given in \cite{M8} in terms of conformal time. Using the
scale factor $a$ as the dynamical variable (as in \cite{M2,M1}),
we get
\begin{equation}
{d^{2}\Phi\over da^{2}}+{1\over 2a}\left(7+6c_{\rm s}^2-
3w\right){d\Phi\over da}+{3\over a^{2}}\left(c_{\rm s}^2-w\right)\Phi
-{1\over a^{4}H^{2}}\nabla^{2}\Phi=0 \,.
\label{20a}\end{equation}
Decomposing $\Phi(a,\vec{x})$ into eigenmodes
$\tilde{\Phi}(a,\vec{k})$ of the comoving Laplacian $\nabla^2$,
and using equations (\ref{7}) and (\ref{8}),
this equation becomes
\begin{eqnarray}
&& a^2{d^{2}\tilde{\Phi}\over da^{2}}
+4a\left({a^2\over a^2+a_{\rm e}^2}\right)
{d\tilde{\Phi}\over da} \nonumber \\
&&{}~~+\left[\pi^2\left({k\over k_{\rm e}}\right)^2
\left({a^{2}+a_{\rm e}^2\over a_{\rm e}a}\right)^2
-2\left({a_{\rm e}^2\over a^{2}+a_{\rm e}^2}\right)\right]\tilde{\Phi}
=0\,.
\label{20}\end{eqnarray}

The evolution of the 
modes $\tilde{h}(a,\vec{k})$ of gauge-invariant tensor perturbations
is given by \cite{M8}
\[
{d^2\tilde{h}\over d\eta^{2}}+2aH{d\tilde{h}\over d\eta}+k^{2}\tilde{h}=0\,.
\]
With $a$ as
the dynamical variable, and using equations (\ref{3}),
(\ref{7}) and (\ref{8}),
this becomes
\begin{equation}
a^2{d^{2}{\tilde{h}}\over da^{2}}+2a\left({a^2+2a_{\rm e}^2
\over a^{2}+a_{\rm e}^2}\right)
{d{\tilde{h}}\over da}+\pi^2\left({k\over k_{\rm e}}\right)^2
\left({a^{2}+a_{\rm e}^2\over 
a_{\rm e}a}\right)^2{\tilde{h}}=0\,.\label{23}
\end{equation}

Super-horizon scales are characterized
by $k\ll aH$, so that by equation (\ref{20a}), we can neglect the
$k$-term in equation (\ref{20}). Similarly, the $k$-term in 
equation (\ref{23}) may be neglected.

\section{Perturbation solutions}

\subsection{Super-horizon perturbations}

For those modes which leave the Hubble radius (necessarily
during inflation), while they remain
outside the Hubble radius, 
we can neglect the $k$ term in
equation (\ref{20}), and use 
the standard transformation for removing the first derivative
\cite{M4}, i.e.
$\phi=\tilde{\Phi}\exp\int 2ada/(a^2+a_{\rm e}^2)$. 
This brings the equation
into the remarkably simple form
\[
a^2{d^2\phi\over da^2}-2\phi=0\,,
\]
which has the explicit exact solution
\[
\phi=C_{1}a^2+C_{2}a^{-1}\,,
\]
where $C_{1}$ and $C_{2}$ 
are arbitrary constants. Thus the exact solution for
super-horizon scalar perturbations (with
$k\ll aH$) is
\begin{equation}
\tilde{\Phi}=A_k\left({a^{2}\over a^{2}+a_{\rm e}^2}\right)+
B_k\left({a_{\rm e}
\over a}\right)\left({a_{\rm e}^2\over a^{2}+
a_{\rm e}^2}\right)\,,
\label{15}\end{equation}
where $A_k$ and $B_k$ are arbitrary dimensionless constants,
the latter corresponding to the decaying modes.
In the vacuum-dominated era, when $a\ll a_{\rm e}$, and
assuming that the scales leave the Hubble radius well before exit,
it follows that $|\tilde\Phi|$ grows as $a^2$. In
the radiation-dominated era ($a\gg a_{\rm e}$),
$|\tilde\Phi|$ is approximately constant (while the scales are still 
super-horizon).
Therefore we have a consistent model in which the 
super-horizon scalar 
perturbations are known exactly via equation (\ref{15}).
They are strongly amplified during inflation and 
then remain approximately constant after inflation. This is
in line with standard results that use an instantaneous
transition \cite{M8},
as well as with the approximate results for
toy-model smooth transitions \cite{M1,g,ms}.
It is also interesting to note that $\Phi=0$ in a
de Sitter model \cite{M8}. Although our model is asymptotically
de Sitter, $\Phi$ quickly grows quadratically through the
almost-de Sitter era.

For adiabatic super-horizon scalar perturbations, the growing
modes have a conserved quantity, given by \cite{l2}
\begin{equation}
\zeta =\tilde{\Phi}+{2\over 3(1+w)}\left[\tilde{\Phi}+
a{d\tilde{\Phi}\over d a}\right]
\,.\label{16}\end{equation}
This is usually used in instantaneous transition models to
express the growing perturbations at late times in terms of their
early-time forms \cite{M8}. In \cite{M1,g}, $\zeta$ is used
to estimate the amplification of perturbations with smooth transition.
We can use our exact solution for a smooth transition,
to show that
in our case, $\zeta$ is exactly constant, even if we include
the decaying modes ($B_k\neq0$). Substituting from (\ref{7}) and
(\ref{15}) into (\ref{16}), we find that (to lowest order in $k$; 
\cite{ms})
\begin{equation}
\zeta={\textstyle{3\over2}}A_{k}\,.\label{18}
\end{equation}
In particular, it follows that for the pure decaying modes ($A_k=0$),
we have $\zeta$ exactly zero. These results about the decaying modes
differ from the standard formulation that $\zeta$ is only conserved
for growing modes, but they are in line with the analysis in \cite{g}.
Of course, the fact that $\zeta=0$ for the decaying modes shows that
$\zeta$ is only useful as a conserved quantity for the growing modes.

For tensor perturbations on scales beyond the Hubble radius,
so that $k\ll aH$,
we can neglect the $k$ term in 
equation (\ref{23}), leading to the exact solution
\begin{equation}
\tilde{h}=C_{k}-D_{k}\left[{a_{\rm e}\over a}+
{1\over 3}\left(a_{\rm e}\over a\right)^{3}\right]\,,
\label{15c}\end{equation}
where $C_{k}$ and $D_{k}$ are arbitrary dimensionless constants, the latter 
corresponding to the decaying modes and so the amplitude of the gravity waves 
is approximately constant.
During inflation, when $a\ll a_{\rm e}$, and
assuming that the scales leave the Hubble radius well before exit, 
equation (\ref{15c}) shows that
\[
\tilde{h}\approx C_{k}-{1\over 3}D_{k}\left({a_{\rm e}\over a}\right)^{3}\,.
\]
During radiation domination ($a\gg a_{\rm e}$), assuming that
the wavelength
is still super-horizon, equation (\ref{15c}) leads to
\[
\tilde{h} \approx C_k-D_k{a_{\rm e}\over a} \,.
\]

\subsection{Sub-horizon perturbations}

For perturbations on scales that are inside the Hubble
radius, i.e. with $k/(aH)$ not negligible,
equations (\ref{20}) and (\ref{23})
can not be solved exactly. However, we can give analytic forms for the
solutions in the vacuum-dominated and radiation-dominated eras,
since the equations reduce to Bessel form. Using \cite{M4,M10}, with
$Z_\nu$ denoting a linear combination of the Bessel functions
$J_\nu$ and $Y_\nu$, 
we find
the following.\\

Vacuum-dominated era:
\begin{eqnarray}
\tilde{\Phi} &\approx& \left({a\over a_{\rm e}}\right)^{1/2}
Z_{-3/2}\left(-\pi{k\over k_{\rm e}}{a_{\rm e}\over a}\right)
\nonumber\\
&=& E_{k}\left({a\over a_{\rm e}}\right)\left[\pi{k\over k_{\rm e}}
\sin\left(\pi{k\over k_{\rm e}}{a_{\rm e}\over a}\right)
+{a\over a_{\rm e}}\cos\left(\pi{k\over k_{\rm e}}{a_{\rm e}\over a}
\right)\right]
\nonumber\\
&&{}+F_{k}\left({a\over a_{\rm e}}\right)
\left[{a\over a_{\rm e}}\sin\left(\pi{k\over k_{\rm e}}{a_{\rm e}
\over a}\right)-
\pi{k\over k_{\rm e}}\cos\left(\pi{k\over k_{\rm e}}{a_{\rm e}\over a}
\right)\right]\,,
\label{s1}    \\
\tilde{h} &\approx & \left({a_{\rm e}\over a}\right)^{3/2}
Z_{-3/2}\left(-\pi{k\over k_{\rm e}}{a_{\rm e}\over a}\right)\nonumber\\
& = & G_{k}
\left[\pi{k\over k_{\rm e}}{a_{\rm e}\over a}\sin
\left(
\pi{k\over k_{\rm e}}{a_{\rm e}\over a}\right)
-\cos
\left(
\pi{k\over k_{\rm e}}{a_{\rm e}\over a}\right)\right]\nonumber\\
&&{}+
H_{k}
\left[\pi{k\over k_{\rm e}}{a_{\rm e}\over a}\cos
\left(
\pi{k\over k_{\rm e}}{a_{\rm e}\over a}\right)
+\sin
\left(
\pi{k\over k_{\rm e}}{a_{\rm e}\over a}\right)\right]
\,. \label{s2}
\end{eqnarray}\\

Radiation-dominated era:
\begin{eqnarray}
\tilde{\Phi} &\approx& \left({a_{\rm e}\over a}\right)^{3/2}
Z_{3/2}\left(\pi{k\over k_{\rm e}}{a\over a_{\rm e}}\right)
\nonumber\\
&=&
I_{k}\left({a_{\rm e}\over a}\right)^{2}
\left[{a_{\rm e}\over a}\sin\left(\pi{k\over k_{\rm e}}{a\over
a_{\rm e}}\right)-
\pi{k\over k_{\rm e}}\cos\left(\pi{k\over k_{\rm e}}{a\over
a_{\rm e}}\right)
\right]\nonumber\\
&&{}+J_{k}\left({a_{\rm e}\over a}\right)^{2}
\left[\pi{k\over k_{\rm e}}\sin\left
(\pi{k\over k_{\rm e}}{a\over a_{\rm e}}\right)+{a_{\rm e}\over a}
\cos\left(\pi{k\over k_{\rm e}}{a\over a_{\rm e}}\right)\right]\,,
\label{s3}\\
\tilde{h} &\approx & \left({a_{\rm e}\over a}\right)^{1/2}
Z_{1/2}\left(\pi{k\over k_{\rm e}}{a\over a_{\rm e}}
\right)\nonumber\\ 
& = &
\left({a_{\rm e}\over a}\right)\left[L_{k}\sin
\left(\pi{k\over k_{\rm e}}{a\over 
a_{\rm e}}\right)
-M_{k}\cos
\left(\pi{k\over k_{\rm e}}{a\over 
a_{\rm e}}\right)\right]
\,.
\label{s4}
\end{eqnarray}

Note that smooth toy models of standard inflation are only applicable
for super-horizon modes around the time of transition,
since on sub-horizon scales, the dynamics of the reheating era have
a significant effect. This is not the case in the decaying vacuum
model, which has no reheating era.

\section{Concluding remarks}

We have shown that a simple decaying vacuum model with a smooth
transition from inflation to radiation domination has
remarkably simple exact forms for its super-horizon
perturbations, and that these forms confirm
recent work on amplification and on the conserved quantity $\zeta$.
However, the model remains simplistic in the sense that it does not
incorporate the second transition, i.e. from radiation to matter
domination. Such an extension is necessary in order to provide
quantitative predictions for the degree of amplification in
modes which affect the microwave background and structure formation,
as well as for the relative contribution of tensor perturbations.

In fact the decaying vacuum model can be extended to incorporate
the creation of massive as well as massless particles, on the basis
of the same physical requirements as discussed above. In this model,
a simple Hubble rate that satisfies the requirements is given
by \cite{n}
\begin{equation}
H(a)=2H_{\rm e}\left[{a_{\rm e}^2\over a^2+a_{\rm e}^2}\right]
\left[{a^2+a_{\rm m}^2 \over a_{\rm m}^{1/2}\left(a^{3/2}
+a_{\rm m}^{3/2}\right)}\right] \,,
\label{s5}\end{equation}
where $a_{\rm e}$ is approximately the epoch of exit (i.e.
$\ddot{a}_{\rm e}\approx 0$), and $a_{\rm m}$ ($\gg a_{\rm e}$)
is approximately the epoch of matter-radiation equality. From
(\ref{s5}) we see that
\begin{eqnarray*}
H &\sim & \mbox{ const }~~\mbox{ for }~~a\ll a_{\rm e} \,, \\
H &\sim & a^{-2}~~\mbox{ for }~~a_{\rm e}\ll a\ll a_{\rm m} \,, \\
H &\sim & a^{-3/2}~~\mbox{ for }~~a_{\rm m}\ll a \,.
\end{eqnarray*}
Thus (\ref{s5}) describes a smooth evolution from inflation to
radiation domination to matter domination, and it can be used
to find the properties of super-horizon perturbations that
leave the Hubble radius during inflation and re-enter soon after
matter-radiation equality. Clearly, it is no longer possible to
find exact analytic forms for these perturbations, and numerical
integration will be necessary. This is the subject of further work.

\end{document}